\documentclass[preprint,showpacs,preprintnumbers,amsmath,amssymb]{revtex4}
\usepackage{graphicx}
\usepackage{epsfig}
\usepackage{lscape}
\usepackage{dcolumn}
\usepackage{bm}
\begin{document}
\title{ Density functional theory calculation of ground state energy, dipole polarizability and hyperpolarizability of a confined helium atom.}
\author{Subhajit Waugh$^{a}$, Avijit Chowdhury$^{b}$, Arup Banerjee$^{a}$,\\
(a) Laser Physics Application Division, Raja Ramanna Centre for Advanced Technology\\
Indore 452013, India\\
(b) BARC Training School at RRCAT, Raja Ramanna Centre for Advanced Technology\\
Indore 452013, India}

\begin{abstract}
We calculate ground-state energies and densities of a helium atom confined in an impenetrable spherical box within density functional theory. These calculations are performed by variationally solving Kohn-Sham equation with the ground-state orbital expanded in terms of Slater-type orbitals. Using the ground-state densities we then calculate static linear polarizability and nonlinear hyperpolarizability and study their variation with the radius of confinement. We find that polarizability decreases monotonically with decreasing confinement radius and the hyperpolarizability not only decreases but also undergoes a change in sign in the strong confinement regime. 
\end{abstract}
\maketitle 
\section{Introduction}
The properties of atoms and molecules undergo drastic change when they are spatially confined in either penetrable or impenetrable cavity as compared to their free counterparts. In recent years the topic of confined atoms has been attracting lot of attention and it has become a field of active research \cite{advancequantumchem}.  
The main reason for this interest in spatially confined atoms and molecules is their applicability to several problems of physics and chemistry. For example, the model of an atom confined in an impenetrable sphere has been employed to simulate the effect of high pressure on the physical properties of atoms, ions, and molecules \cite{michels,sommerfield,tenseldam}. The study of confined atoms  provides insight into various properties of quantum nanostructures like quantum dots or artificial atoms \cite{sako1,sako2}. For more detailed discussion on these applications we refer the reader to review articles \cite{dolmatov,connerade,jaskolski}. 

Recently numerous studies on helium atom confined in an impenetrable spherical cavity \cite{tenseldam,gimarc,ludena1,ludena2,joslin,aquino1,patil,banerjee1,aquino2,flores1,laughlin,flores2} and  in a penetrable spherical cavity \cite{martin1,martin2} as well have been reported in the literature. Helium atom being the simplest many-electron system, the confined version of this atom provides a lucid way to study the effect of confinement on the electron correlation which arises due to coulomb interaction between the two electrons and pauli exclusion principle. Besides helium atom, few studies on some more confined many-electron atoms up to neon atom have also been reported in the literature \cite{ludena1,faassen,diaz}. We note here that most of the studies on confined helium atom mainly considered the ground-state electronic properties. In contrast not many studies have been carried out on the electric response properties like dipole polarizability and hyperpolarizablity of this system. To the best of our knowledge results for the linear dipole polarizability of confined helium atom were reported only in Refs. \cite{tenseldam,faassen}. However, no study devoted to the calculation of hyperpolarizability of confined helium atom exists in the literature. The main aim of this paper is to carry out calculation of not only linear dipole polarizability but also nonlinear hyperpolarizability of confined helium atom and study the evolution of these quantities with the size of the cavity (or strength of the confinement). 
We wish to point out here that the calculation of both linear polarizabilty ($\alpha$) and third-order hyperpolarizability ($\gamma$) of confined hydrogen atom were carried out in Ref. \cite{banerjee2}. It was shown that both $\alpha$ and $\gamma$ strongly depend upon the radius of confinement and moreover, $\gamma$ changes sign and becomes negative under strong confinement. Thus it will be interesting to find out when $\gamma$ of confined helium atom undergo reversal of sign. 

In this work we carry out calculation of electric response properties by employing density functional theory (DFT) based
variation-perturbation approach \cite{banerjee3}. To carry out these calculations we need to have ground-state densities of confined helium atom. This task has been accomplished by employing a variational approach involving minimization of the ground-state energy fuctional within the realm of DFT. A brief description of the theoretical methods employed for calculations of both ground-state densities and electrical response properties are presented in Section 2. The Section 3 is devoted to the discussion of results and paper is concluded in Section 4 
 
\section{Method of Calculations}
We begin this section with a brief description of the method for obtatining ground-state density of confined helium atom within the realm of DFT. The Kohn-Sham (KS) equation of DFT are obtained by minimizing the energy functional (in atomic units) \cite{parr}
\begin{equation}
E_{KS} [\rho ] = T_{s}[\rho ] + J[\rho ] + E_{xc}[\rho ] + \int v_{ext}({\bf r})\rho ({\bf r})d{\bf r}.
\label{energyks}
\end{equation}
Here $T_{s}[\rho ]$ denote the kinetic energy functional of non-interacting particles and in terms of single-particle orbitals $\psi_{i}({\bf r})$ it is represented as
\begin{equation}
T_{s}[\rho ]  =  \sum_{i=1}^{N}\int \psi^{*}_{i}({\bf r})(-\frac{1}{2}\nabla^{2})
\psi_{i}({\bf r})d{\bf r}.        
\end{equation}  
These orbitals yield density of interacting system via 
\begin{equation}
\rho({\bf r}) = \sum_{i=1}^{N}\sum_{s}|\psi_{i}({\bf r},s)|^{2}.
\label{rho}
\end{equation}
In Eq. (\ref{energyks}) J$[\rho ]$ represent the classical part of the electron-electron repulsion, E$_{xc}[\rho ]$ is the exchange-correlation functional and the last term corresponds to the contribution due to the external potential $v_{ext}({\bf r})$. For confined atom $v_{ext}({\bf r})$ has two parts namely: (1) the nuclear potential $-Z/r$ (where Z is the nuclear charge) and (2) the confining potential $v_{conf}$ due to an impentrable spherical box of radius $r_{c}$ of the form 
\begin{equation}
v_{conf}({\bf r}) =
\left\{\begin{array}{cl}
0 & \mbox{$r < r_{c}$} \\
\infty & \mbox{$r \geq r_{c}$}
\end{array}\right.
\label{cofpot}
\end{equation}
The minimization of E$_{KS}[\rho ]$ with respect to single-particle orbital $\psi_{i}({\bf r})$ satisfying the condition given by Eq. (\ref{rho}) leads to so-called Kohn-Sham equation which is the workhorse of DFT.
In this paper we carry out this minimization explicitly by using appropriate variational forms for the single-particle orbitals. For this purpose we expand the single-particle orbital of a confined helium atom as
\begin{equation}
\psi_{1s}({\bf r}) = \sum_{i}c_{i}\chi_{i}({\bf r})f_{c}(r)
\end{equation}
where $\{c_{i}\}$ are the variational parameters which are determined by minimization of the ground-state energy (Eq. (\ref{energyks})), f$_{c}(r)$ is the cut-off function which takes care of the confinement boundary condition of density vanishing at $r = r_{c}$, and the basis function $\chi_{i}({\bf r})$ is given by the product of a Slater-type orbital (STO) for the radial part and a spherical harmonic function Y$_{lm}(\theta,\phi)$ for the angular part as
\begin{equation}
\chi({\bf r}) = R_{nl}(r)Y_{lm}(\theta,\phi).
\end{equation}
The radial function R$_{nl}(r)$ is given by
\begin{equation}
R_{nl}(r) = r^{n-1}e^{-\zeta r}
\end{equation}
with $n$ and $\zeta$ representing orbital parameters which we choose from Ref. \cite{ludena1}. For cut-off function f$_{c}(r)$ we choose both linear
\begin{equation}
f_{c}(r) = \left (1 - \frac{r}{r_{c}}\right )
\label{lincutoff}
\end{equation}
and the quadratic 
\begin{equation}
f_{c}(r) = \left (1 - \frac{r^{2}}{r_{c}^{2}}\right )
\label{quadcutoff}
\end{equation}
forms and investigate their performance in yielding ground-state energies of confined helium atom.  

Having described the variational approach for obtaining ground-state density of confined helium atom we next briefly  outline the method adopted in this paper to calculate static linear ($\alpha$) and nonlinear ($\gamma$) polarizabilities. The response properties mentioned above are calculated by employing variation-perturbation (VP) approach within DFT. In density based VP, energy to order
(2n + 1) is determined by the perturbation expansion of the density correct to order $n$ only. 
Further, the even-order energy correction E$^{(2n+2)}$ is minimum for the exact induced 
density $\rho^{(n+1)}$, if expansion up to order $n$ is known
exactly. For details of the VP approach within DFT, we refer the reader to
reference \cite{banerjee3}. For our purpose here it is sufficient to note that $\alpha$ and $\gamma$
are calculated from the second-order  $\Delta E^{(2)}$ and the fourth-order  $\Delta E^{(4)}$ change in energies
respectively, by employing relations 
\begin{eqnarray}
\alpha & = & -2\Delta E^{(2)} \nonumber \\
\gamma & = & -24\Delta E^{(4)} 
\end{eqnarray}
These energy changes are in turn obtained variationally by minimising  
\begin{equation}
E^{(2)} = \int v^{(1)}({\bf r_1})\rho^{(1)}({\bf r_1})d{\bf r_1} + 
\frac{1}{2}\int\frac{\delta^{(2)}F[\rho_{0}]}{\delta\rho({\bf r}_1)
\delta\rho({\bf r}_2)}\rho^{(1)}({\bf r}_1)
\rho^{(1)}({\bf r}_2)d{\bf r}_{1}d{\bf r}_{2},
\label{eq2.15}
\end{equation}
with respect to $\rho^{(1)}$, and 
\begin{eqnarray}
E^{(4)} & = &  \frac{1}{2}\int\frac{\delta
^{(2)}F[\rho_{0}]}{\delta\rho({\bf r}_1)\delta\rho({\bf r}_2)}
\rho^{(2)}({\bf r}_1)\rho^{(2)}({\bf r}_2)d{\bf r}_{1}d{\bf r}_{2} \nonumber \\
        & +  &    \frac{1}{2}\int\frac{\delta
^{(3)}F[\rho_{0}]}{\delta\rho({\bf r}_1)\delta\rho({\bf r}_2)
\delta\rho({\bf r}_3)}\rho^{(1)}({\bf r}_1)
\rho^{(1)}({\bf r}_2)\rho^{(2)}({\bf r}_3)d{\bf r}_{1}d{\bf r}_{2}d{\bf r}_{3}
\nonumber \\
        & +  &    \frac{1}{24}\int\frac{\delta
^{(4)}F[\rho_{0}]}{\delta\rho({\bf r}_1)\delta\rho({\bf r}_2)
\delta\rho({\bf r}_3)\delta\rho({\bf r}_4)}\rho^{(1)}({\bf r}_1)
\rho^{(1)}({\bf r}_2)\rho^{(1)}({\bf r}_3)\rho^{(1)}({\bf r}_4)\nonumber \\
        &\times & d{\bf r}_{1}d{\bf r}_{2}d{\bf r}_{3}d{\bf r}_{4}.
\label{eq2.25}
\end{eqnarray}
with respect to $\rho^{(2)}$.
Here v$^{(1)}({\bf r})$ is the applied (external) perturbation. F$[\rho]$ is a universal functional of the density and it is given by the sum of the kinetic, 
Hartree and the exchange-correlation energies of the electrons. All the functional derivative in the equations
above (Eqs.(\ref{eq2.15}) and (\ref{eq2.25})) are evaluated at the ground-state density $\rho_{0}$. For an atom placed in a static electric field ${\cal E}$ along z-axis the variational ansatz for $\rho^{(1)}$ and $\rho^{(2)}$ are
\begin{eqnarray}
\rho^{(1)}({\bf r}) & = & \Delta_{1}(r)cos\theta\rho_{0}(r), \nonumber \\
\rho^{(2)}({\bf r}) & = &  [\Delta_{2}(r) + \Delta_{3}(r)cos^{2}\theta]
\rho_{0}(r) + \lambda\rho^{(0)}({\bf r})
\label{eq2.48}
\end{eqnarray}
where
\begin{equation}
\Delta_{i}(r) = a_{i}r + b_{i}r^{2} + c_{i}r^{3} + \cdots ,i = 1,2,3 \cdots
\label{eq2.49}
\end{equation}
with a$_{i}$, b$_{i}$ $\cdots$ being the variationals parameters. $\lambda$
is fixed for each set of parameters by the second-order normalization condition
$\int\rho^{(2)}({\bf r})d{\bf r} = 0$. Notice that the first-order 
normalization condition $\int\rho^{(1)}({\bf r})d{\bf r} = 0$  is 
automatically satisfied by $\rho^{(1)}({\bf r})$ in Eq.(\ref{eq2.48}). We have
used five parameters for $\Delta_{1}$ and eight parameters each for
$\Delta_{2}$ and $\Delta_{3}$. Adding more parameters does not affect the 
results significantly indicating their convergence. For evaluating the functional derivatives of 
the the exchange and correlation energies, we use the Dirac functional \cite{dirac} for the exchange contribution and Gunnarsson-Lundquist (GL) parametrization \cite{gl} for the correlation energy within local-density approximation (LDA). In the next section we discuss results obtained by us using above-mentioned methods.

\section{Results and Discussions}  
We begin this section with the discussions of the results for ground-state energy of a confined helium atom obtained by us to assess their accuracies. In this connection we note that DFT based results for confined helium atom have already been reported in Ref. \cite{aquino2} which were obtained by numerically integrating the KS equation with Dirichilet boundary condition \cite{garza}. In order to establish the accuracy of our variational results we compare them with those of Ref. \cite{aquino2}. First of all we note that we perform calcualtion with both linear and quadartic cut-off functions as given by Eqs. (\ref{lincutoff}) and (\ref{quadcutoff}) respectively. We find that the ground-state energies for several values of confinement radius $r_{c}$ obtained with quadratic cut-off function are close but slightly lower than the corresponding results obtained with linear cut-off functions. Therefore, in the following we report results only with quadratic cut-off functions. 

In table I we present the results for the energies for the ground-state $^{1}S (1s^{2})$  of a confined helium atom as a function of confinement radius $r_{c}$. In this table we present the results for the case of exact exchange (EXX) energy, which are obtained by substituting $E_{X} = -E_{H}/2$ (exact for two-electron systems) and E$_{C} = 0$. This case corresponds to Hatree-Fock (HF) approximation and we compare our EXX results with those of Ref. \cite{ludena1}. We also present the results obtained with exchange-only (XO) with E$_{C} = 0$ and exchange-correlation (XC) energies within LDA in second and third columns respectively. These results are compared with the corresponding numbers of Ref. \cite{aquino2} which were obtained by numerically solving the Konn-Sham equation with Dirchilet boundary condition. In order to assess the accuracy of DFT based results we also display results obtained via correlated wavefunction based calculation with 7-parameter Hylleraas expansion \cite{flores1} in the last column of Table I. First we note that the results for the case of EXX energy obtained by us match very well up to 4-th decimal place with the results of Ref. \cite{ludena1} for all values of $r_{c}$. This establishes the accuracy of the variational method employed by us. The XO-LDA results obtained by us are close but slightly higer than the corresponding EXX values as long as  $ r_{c} \leq 1.0$ $a. u.$. On the other hand, for $ r_{c} > 1.5$ $a. u.$ we find that trend is just reverse. Moreover, XO-LDA results are also quite close to the data available for the range $r_{c} = 2 - 6$ $a.u.$ in Ref. \cite{aquino2}. With the inclusion of correlation energy term within LDA the ground-state energies of confined helium atom reduce slightly as compared to the corresponding XO numbers. Our XC results match well with those of Ref. \cite{aquino2} and the difference in the two results are mainly due to the use of two different XC functionals for the calculations. We note here that in Ref. \cite{aquino2} Pewrdew-Wang form of the correlation functional \cite{perdewang} along with Dirac form for the exchange energy functional has been employed whereas we employ GL parametrization for the correlation part \cite{gl}. With the inclusion of correlation, the ground-state energy of a confined helium atom decreases relative to the corresponding XO-LDA valuses as long as energies remain positive. For confinement radii with negative ground-state energies inclusion of correlation leads to lowering of the ground-state energy. Similar trend is also observed with Perwdew-Wang XC functional.  

The comparison of DFT based results with  the corresponding Hylleraas wavefunction based numbers clearly shows the EXX results are the closest to the latter.  From these results we conclude that the contribution of correlation energy both in strong ($r_{c} \leq 1$ $a. u. $) and weak ($r_{c} \geq 1$ $a. u. $) confinemnets is not very significant to the total energy of a confined helium atom. Thus for confined helium atom it is possible to get sufficiently accurate results for the ground-state energy provided exchange part of the energy is accurately taken into account.  

Having established the accuracy of the DFT based results for the ground-state energy of a confined helium atom we now present the results for static polarizability $\alpha$ and second-order hyperpolarizabilty $\gamma$ of confined helium atom by employing the ground-state densities which are obtained via above-mentioned variational ground-state energy calculations. In Table II we compile the results for both $\alpha$ and $\gamma$ corresponding to several values of  $r_{c}$ obtained with  EXX, XO-LDA and XC-LDA ground-state densities. It can be seen from Table II that for all the three energy functionals $\alpha$ decreases monotonically with decrease in $r_{c}$ and appproaches zero for very small value of $r_{c}$. This trend is in agreement with the results of Ref. \cite{faassen}. We also note that with increase in  $r_{c}$ the values of $\alpha$ correctly tend to the respective free atom cases which are $\alpha_{EXX} = 1.33$ $a. u. $, $\alpha_{XO} = 1.77$ $a. u.$, and  $\alpha_{XC} = 1.63$ $a. u.$ \cite{banerjee4}. It is also interesting to note from Table II that for $r_{c}<1.5$ $a. u.$ the values of polarizability $\alpha$ obtained with three differnt densities are almost identical. Therefore, like ground-state energy of a confined helium atom its polarizability too does not have a strong dependence on the correlation energy specially in the strong confinment regime. Next we focus our attention on the results for the hyperpolarizability $\gamma$ of a confined helium atom, which are to our knowledge not reported earlier. The results for $\gamma$ also correctly converge to their respective free atom values with the increase in $r_{c}$. These values are $\gamma_{EXX} = 36.2$ $a. u. $, $\gamma_{XO} = 114.2$ $ a. u. $, and  $\gamma_{XC} = 88.15$ $a. u.$ \cite{banerjee4}. Like polarizability, the values of $\gamma$ also show a monotonic decreasing trend, however, the decrease in the values of $\gamma$ are more rapid as comapred to the polarizablity. With the increase in the compression, that is decrease in $r_{c}$ the values of hyperpolarizability $\gamma$ not only approach zero but also undergo a change in sign. The change in sign occurs at different value of $r_{c}$ for three different ground-state densities employed in this paper. The change in the sign of hyperpolarizabilty of a strongly confined hydrogen atom has already been discussed in Ref. \cite{banerjee2}. We note here that the change in sign of $\gamma$ is akin to the term \textit{hypopolarizability} which was coined by Coulson et al. \cite{coulson}. The change in sign of $\gamma$ was later discussed by Langhoff et al. \cite{langhoff} in connection with study of hyperpolarizability of high Z ions isoelectronic with Na and Mg series. They concluded that the compact electronic charge density resulting from the increasing value of Z within isoelectonic series is responsible for reversal of sign of $\gamma$. Similarly in a confined atom the charge density becomes highly compact with decreasing radius of confinement thereby yielding a negative hyperpolarizability. Finally, we note that unlike polarizabilty the results for $\gamma$ obtained with different densities vary for all values of $r_{c}$. This indicates that hyperpolarizability $\gamma$ depends crucially on the nature of ground-state density employed for the calculation.  

\section{Conclusion}
In this paper we have calculated the ground-state energies and the densities of a helium atom confined in an impentrable spherical box for various values of radius of confinement within DFT.  We use three different exchange-correlation energies namely, exact exchange equivalent to HF level, exchange-only at LDA level, and correlation included with exchange at LDA level to obtain the ground-state properties. Using these ground-state densities we perform calculations of static linear and nonlinear electric response propertise of confined helium atom and study their variation with the strength of the confinment. The ground-state energies and densities are obtained by variationally solving the KS equation of DFT with the variational form for the ground-state orbital expanded in terms of STOs. The results obtained by us are quite accurate and compare well with the already published data. We find that energies of a confined helium atom are not affected by the inclusion of correlation energy especially in the strong confiment regimes. With the increase in the value of radius of confinement the differences between the results for energy obtained by employing three different densities increase indicating the importance of correlation energy for determining the energy of a free helium atom. The linear polarizability $\alpha$ of a confined helium atom decreases with the decrease in the value of confinement radius approacing the value of zero in the strong confinement regime. Results of our calculations clearly demonstrate that for $r_{c} <1.5$ $a. u. $ the values of $\alpha$ do not depend much on the correlation energy. On the other hand, value of hyperpolarizability $\gamma$  shows a decreasing trend with decreasing $r_{c}$ which approaches the value of zero rapidly and then changes sign on further decreasing $r_{c}$. This change in sign of $\gamma$ is attributed to charge density becoming highly compact in the case of strong confinement.

\acknowledgments{ We dedicate this paper to Prof. K. D. Sen who introduced us to the field of confined atoms. S. W. and A. B. wish to thank Dr. S. C. Mehendale for his constant support}   
 
\newpage
\begin{table}
\caption{Ground-state energies of confined helium atom as a function of confinment radius $r_{c}$ (in atomic units) for the case of exact exchange (EXX), exchange-only within LDA (XO-LDA) and exchange-correlation within LDA (XC-LDA). The numbers in the parenthesis are taken from already published data. The results in the last column are taken from \cite{flores1}}
\begin{center}
\begin{tabular}{c|c|c|c|c}\hline
$r_{c}$ & EXX  & XO-LDA & XC-LDA & Correlated-Hylleraas \\
        &(Ref.\cite{ludena1})&  (Ref. \cite{aquino2})& (Ref. \cite{aquino2})  & (Ref. \cite{flores1})  \\
\hline
0.5 & 22.79096 & 23.32202 & 23.099 & 22.7419 \\
    & (22.79095) & - &  - & \\
0.6 & 13.36683  & 13.81792 &  13.605 & 13.3187 \\
    &  (13.36682)  & - &  -  &     \\ 
0.7 &  7.97302  &  8.36716 & 8.164 & 7.9258 \\
    &  (7.97302) &  - &  -    &  \\ 
0.8 &  4.65736  &  5.00895  & 4.8133 & 4.6110 \\
    &  (4.65737) & - & -    &          \\ 
0.9 &  2.50942  & 2.82805  & 2.369 & 2.4638 \\
    &  (2.50944) & - &  -     &      \\               
1.0 & 1.06120 & 1.35362 & 1.17040&  1.063 \\
    & (1.01624) &  -    &   -   &       \\
1.5 & -1.86422 & -1.64897 &-1.81185 & -1.9066 \\
    & (-1.86422)  & -     &   -  &    \\
2.0 & -2.56257 & -2.39363 & -2.53480 &  -2.6035  \\
    &  (-2.56253)& (-2.38363)& (-2.50589) &      \\ 
3.0 & -2.83078 & -2.68201 &-2.82256 & -2.8715\\
    & (-2.83083) & (-2.68210) & (-2.79608) &  \\
4.0 & -2.85854 & -2.71807 & -2.85552 & -2.8994 \\
    &  (-2.85852) &(-2.71813) & (-2.82970)&   \\
5.0 & -2.86138 & -2.72288 & -2.85856 & -2.9026\\
    &  (-2.86134)& (-2.72290)& (-2.83387)&    \\
6.0& -2.86162 &-2.72346  & -2.86000 & -2.9032 \\
   &  (-2.86151)&(-2.72354)& (-2.83439) &    \\
\hline
\end{tabular}
\end{center}
\end{table}

\begin{table}
\caption{Static polarizability $\alpha$ and hyperpolarizability $\gamma$ of a confined helium atom for various values of confinemnt radius $r_{c}$. All numbers are in atomic unit }
\tabcolsep=0.2in
\begin{tabular}{@{}|c|c|c|c|c|c|c|@{}}\toprule
 & \multicolumn{2}{c|}{EXX} &\multicolumn{2}{c|}{XO-LDA}& \multicolumn{2}{c|}{XC-LDA}  \\
\cline{2-7}
$r_{c}$ & $\alpha$ & $\gamma$& $\alpha$ & $\gamma$ & $\alpha$ & $\gamma$ \\ \hline
1.0 & 0.043 & -1.05$\times10^{-4}$ & 0.043 & -1.16$\times10^{-4}$& 0.043 & -8.78$\times10^{-5}$ \\
1.1 & 0.059 & -1.72$\times10^{-4}$ & 0.060 & -2.04$\times10^{-4}$ & 0.060 & -1.36$\times10^{-4}$ \\
1.2  & 0.079 & -2.15$\times10^{-4}$& 0.080 & -2.97$\times10^{-4}$ & 0.080 &  -1.46$\times10^{-4}$  \\  
1.3  & 0.102&  -1.24$\times10^{-4}$& 0.103 & -3.12$\times10^{-4}$ &  0.103 & -2.08$\times10^{-6}$  \\
1.4  & 0.129  & 3.54$\times10^{-3}$& 0.131 & -7.24$\times10^{-5}$ & 0.131 & 5.22$\times10^{-4}$\\
1.5  & 0.160  & 1.56$\times10^{-3}$ & 0.162 & 7.74$\times10^{-4}$&  0.162 &  1.87$\times10^{-3}$ \\
2.0  &   0.360    &   0.0603 & 0.374 & 0.051 & 0.372 & 0.0634 \\ 
2.5  & 0.606      &  0.487  & 0.647 & 0.469 &0.641  & 0.530  \\
3.0  &   0.839    &  1.900   &0.934 & 2.158 &  0.916&  2.324 \\
4.0  & 1.161      &  10.358  & 1.392 & 14.862 &  1.336 & 14.674   \\
5.0  & 1.286 & 23.381 &1.641 & 45.029  & 1.544& 40.552 \\
6.0  &  1.322   &35.766 & 1.763 & 93.056 & 1.628 & 75.265 \\ 
\hline
\end{tabular}
\end{table}



\clearpage
\newpage 
\begin{thebibliography}{25}
\bibitem{advancequantumchem}S. A. Cruz, Ed. The Theory of Confined Quantum Systems Advances in Quantum Chemistry 57 (2009)  and refeerences therein.
\bibitem{michels} A. Michels, J. de Boer, and A. Bijl, Physica (Amsterdam) 4 (1937) 981.
\bibitem{sommerfield}A. Sommefeld and H. Welker,  Ann. Phys., Lpz. 32 (1938) 56.
\bibitem{tenseldam}C. A. Ten Seldam and S. R. de Groot, Physica 18 (1952) 904.
\bibitem{sako1}T. Sako and G. H. F. Diercksen, J. Phys. B At. Mol. Opt. Phys. 36 (2003) 1433 .
\bibitem{sako2}T. Sako and G. H. F. Diercksen, J. Phys. B At. Mol. Opt. Phys. 36 (2003) 1681 .
\bibitem{dolmatov} V. K. Dolmatov, A. S. Baltenkov, J. -P. Connerade, and S. Manson, Radiation Physics and Chemistry 70(2004) 417  and references theirin.
\bibitem{connerade} J. -P. Connerade and P. Kengkan, Proc. Idea-Finding Symp. Frankfurt Institute for Advanced Studies, (2003) p. 35.
\bibitem{jaskolski} W. Jaskolski, Phys. Rep. 271 (1996) 1 .
\bibitem{gimarc}B. M. Gimarc, J. Chem. Phys. 47 (1967) 5110.
\bibitem{ludena1}E. V. Ludena, J. Chem. Phys. 69 (1978) 1170.
\bibitem{ludena2}E. V. Ludena and M. Greogri, J. Chem. Phys. 71 (1979) 2235.
\bibitem{joslin}C. Joslin and S. Goldman, J. Phys. B: At. Mol. Opt. Phys. 25 (1992) 1965.
\bibitem{aquino1}N. Aquino, A. F-Riveros, J. F. Rivas-Silva, Phys. Lett. A  307 (2003) 326.
\bibitem{patil} S. H. Patil and Y. P. Varshni, Can. J. Phys./Rev. Can. Phys. 82 (2004) 647.
\bibitem{banerjee1}A. Banerjee, C. Kamal, and A. Chowdhury, Phys. Lett. A 350 (2006) 121 . 
\bibitem{aquino2}N. Aquino, J. Garza, A. Flores-Riveros, J. F. Rivas-Silva, and K. D. Sen, J. Chem. Phys. 124 (2006) 054311.
\bibitem{flores1}A. Flores-Riveros, N. Aquino, H. E. Montgomery Jr., Phys. Lett. A 374 (2010) 1246 .
\bibitem{laughlin}C. Laughlin and S. I. Chu, J. Phys. A: Math. Theor. 42 (2009) 265004. 
\bibitem{flores2}H. E. Montgomery Jr, N. Aquino, and A. Flores-Riveros, Phys. Lett. A 374 (2010) 2044.
\bibitem{martin1}J. L. Martin and S. A. Cruz, J. Phys. B: At. Mol. Opt. Phys. 224 (1991) 2899.
\bibitem{martin2}J. L. Martin and S. A. Cruz, J. Phys. B: At. Mol. Opt. Phys. 25 (1992) 4365.
\bibitem{faassen}M. van Faassen, J. Chem Phys 131 (2009) 104108.
\bibitem{diaz}C. Diaz-Garcia and S. A. Cruz, Int. J. Quant. Chem. 108 (2008) 1572.
\bibitem{banerjee2}A. Banerjee, K. D. Sen, J. Garza, and R. Vargas, J. Chem. Phys. 116 (2002) 4054.
\bibitem{banerjee3}M. K. Harbola and A. Banerjee, Phys. Lett. A 222 (1996) 315; A. Banerjee and M.K. Harbola, Pramana J. Phys. 49 (1997) 455; A. Banerjee and M.K. Harbola, Eur. Phys. J. D 1 (1998) 265.
\bibitem{parr}R.G. Parr and W. Yang, Density Functional Theory of
Atoms and Molecules, Oxford University Press, New York, 1989.
\bibitem{dirac}P. A. M. Dirac, Proc. Camb. Phil. Soc. 26 (1930) 376.
\bibitem{gl} O. Gunnarsson and B. I. Lundqvist, Phys. Rev. B 13 (1976) 4274.
\bibitem{garza}J. Garza, R. Vargas, and A. Vela, Phys. Rev. E (1998) 3949.
\bibitem{perdewang}J. P. Perdew and Y. Wang, Phys. Rev. B 45 (1992) 13244.
\bibitem{banerjee4} A. Banerjee and M. K. Harbola, Phys. Rev. A 60 (1999) 3599.
\bibitem{coulson} C. A. Coulson, A. Maccoll, and L. E. Sutton, Trans . Faraday soc. 61 (1952) 106.
\bibitem{langhoff} P. W. Langhoff, J. D. Lyons, and R. P. Hurst, Phys. Rev. 148 (1966) 18.
\end{thebibliography}
\end{document}